\documentclass[%
 reprint,
superscriptaddress,
 amsmath,amssymb,
 aps,
floatfix,
]{revtex4-2}

\usepackage{graphicx}
\usepackage{dcolumn}
\usepackage{bm}
\usepackage{xcolor}
\usepackage{soul}

\usepackage[symbol]{footmisc}

\begin{document}

\preprint{APS/123-QED}

\title{Phase transition kinetics revealed in laser-heated dynamic diamond anvil cells}

\author{Matthew Ricks}
\affiliation{Department of Physics and Astronomy, Brigham Young University, Provo, Utah 84602, USA}

\author{Arianna E. Gleason}
\affiliation{SLAC National Accelerator Laboratory, 2575 Sand Hill Rd., Menlo Park, CA, 94025, USA}

\author{Francesca Miozzi}
\affiliation{Earth and Planets Laboratory, Carnegie Institution for Science, Washington, DC 20015}

\author{Hong Yang}
\affiliation{Department of Earth and Planetary Sciences, Stanford University, 450 Jane Stanford Way,  Building 320, Stanford, CA 94305}

\author{Stella Chariton}
\affiliation{Center for Advanced Radiation Sources, The University of Chicago, Chicago, IL 60637}

\author{Vitali B. Prakapenka}
\affiliation{Center for Advanced Radiation Sources, The University of Chicago, Chicago, IL 60637}

\author{Stanislav V. Sinogeikin}
\affiliation{DAC Tools, LLC, Naperville, Illinois 60565, USA}

\author{Richard L. Sandberg}
\affiliation{Department of Physics and Astronomy, Brigham Young University, Provo, Utah 84602, USA}

\author{Wendy L. Mao}
\affiliation{Department of Earth and Planetary Sciences, Stanford University, 450 Jane Stanford Way,  Building 320, Stanford, CA 94305}

\author{Silvia Pandolfi}
\affiliation{SLAC National Accelerator Laboratory, 2575 Sand Hill Rd., Menlo Park, CA, 94025, USA}
\affiliation{Now at: Sorbonne Université, Muséum National d’Histoire Naturelle, UMR CNRS 7590, Institut de Minéralogie, de Physique des Matériaux et de Cosmochimie, IMPMC, 75005 Paris, France}

\date{\today}

\begin{abstract}
We report on a novel approach to dynamic compression of materials that bridges the gap between previous static- and dynamic-
compression techniques, allowing to explore a wide range of pathways in the pressure-temperature space. By combining a dynamic-diamond anvil cell setup with double-sided laser-heating and in situ X-ray diffraction, we are able to perform dynamic compression at high temperature and characterize structural transitions with unprecedented time resolution. Using this method, we investigate the $\gamma-\epsilon$ phase transition of iron under dynamic compression for the first time, reaching compression rates of hundreds of GPa/s and temperatures of 2000 K. Our results demonstrate a distinct response of the $\gamma-\epsilon$ and $\alpha-\epsilon$ transitions to the high compression rates achieved. These findings open up new avenues to study tailored dynamic compression pathways in the pressure-temperature space and highlight the potential of this platform to capture kinetic effects in a diamond anvil cell.
\end{abstract}

\maketitle

Understanding how matter responds and deforms under high strain rates is crucial for various fields, with applications ranging from planetary formation modeling \cite{MONTEUX2019238}, ballistics \cite{CHAGAS20221799}, and high-strength ceramic development \cite{Godec.2019}. Dynamic compression experiments mostly rely on the generation and propagation of ultrasonic shock waves to reach extreme pressure and temperature (P-T) conditions. Single-shock experiments, however, are constrained to probe states along the Principal Hugoniot of a material \cite{Duffy2019}. Despite recent developments allowing  access to quasi-isentropic pathways using ramp compression, exploring arbitrary P-T conditions via dynamic compression remains a significant challenge and requires precise knowledge of the material's equation of state (EOS) \cite{Dlot2011}. Here, we use a novel dynamic-diamond anvil cell (dDAC) setup compatible with \textit{in situ} laser-heating (LH). This setup enables isothermal compression at high temperature (HT), significantly expanding the range of conditions that can be studied dynamically. We use \textit{in situ} X-ray diffraction (XRD) to study the phase transitions of iron (Fe), and demonstrate compression rates of hundreds of GPa/s at temperatures up to 2000 K. By accessing this largely unexplored region of strain rates, we are able to bridge the gap between previous static- and dynamic-compression techniques, providing valuable experimental benchmarks for deformation and strength models that were still lacking in this regime.

Fe is the main constituent of the Earth's core and a model structural component for engineering applications. As such, its behavior at extreme conditions has been extensively studied, both experimentally and theoretically. The stable structure of Fe at ambient conditions, the so-called $\alpha$ phase with body-centered cubic (\textit{bcc}) structure, transforms into an hexagonal-close packed (\textit{hcp}) structure, the $\epsilon$ phase, under high-pressure (HP) \cite{Shen1998}. The HT phase, $\gamma$, has a face-centered cubic (\textit{fcc}) structure. Upon compression, this phase transforms as well into the $\epsilon$ phase, which remains stable up to multi-Mbar pressures and is believed to be present in the Earth's solid core \cite{anderson1986}. Numerous static compression experiments have investigated Fe behavior at HP-HT \cite{Boehler1993,Shen1998,Komabayashi2009,Kubo2003,Ma2004,Funamori1996, Miozzi2020, Bancroft1956}. To characterize the equilibrium phase diagram and reproduce more accurately the conditions in Earth's interior, efforts were made in these studies to maintain quasi-hydrostatic conditions. On the other hand, dynamic compression techniques, such as gas gun and laser ablation, have been used to characterize Fe deformation and melting at ultrafast timescales, from $\mu$s down to ps \cite{Yoo1993,Benuzzi2002,Ping2013,Amadou2015,Harmand2015,Denoeud2016,Torchio2016,Merkel2021,Branch2018, Boettger1997, Hawreliak2008}. These techniques can attain pressures up to several TPa \cite{Smith2014} (\textit{i.e.,} tens of millions atmospheres); however, they are constrained to probe specific thermodynamic pathways in the P-T space. The Principal Hugoniot of Fe crosses the  $\alpha-\epsilon$ phase boundary, and the transition has been observed at a slightly higher pressure than the value from static compression experiments \cite{Hawreliak2006,Kalantar2005,Hawreliak2021,Hawr2022}. Despite recent developments that allow us to deploy more complex, off-Hugoniot compression profiles \cite{Wang2013,Smith2018,Kraus2022}, the $\gamma-\epsilon$ transition remains inaccessible using conventional dynamic compression techniques.

Recent developments in dDAC technology have started to fill the gap between static- and shock-compression experiments, allowing a detailed study of strain-rate dependendent phenomena. Using either gas supplied membrane or electromechanical piezoelectric actuators to control pressure, dDACs can access compression rates up to 160 TPa/s \cite{sinogeikin2015}. Kon\^opkov\'a \textit{et al.} investigated the  $\alpha-\epsilon$ transition of Fe at compression rates up to 4.1 GPa/s, and found a transition onset consistent with the quasi-equilibrium phase diagram when experiments were conducted in quasi-hydrostatic conditions \cite{Kon2015}. Here, we use a double-sided LH-compatible dDAC that enables simultaneous heating and compression. With this capability, we provide the first insight on the $\gamma-\epsilon$ phase transition of Fe under dynamic compression. The phase transition onset measured for two different temperatures agrees with previous static compression experiments from Shen \textit{et al.} \cite{Shen1998}, while it shows a phase transition lowering with respect to results from more recent studies \cite{Komabayashi2009, Funamori1996, Kubo2003}. The $\alpha-\epsilon$ phase transition was also characterized at ambient temperature. For this transformation, an increase of the transition onset is observed when increasing the compression rate, revealing that the response to high strain rates depends on the specifics of the transition mechanism. Our findings demonstrate the first dDAC platform compatible with laser heating HT experiments, which enables \textit{in situ} characterization of structural transformations over millisecond timescales at varying compression rates and temperatures.
\begin{figure}
\includegraphics[width=0.9\columnwidth]{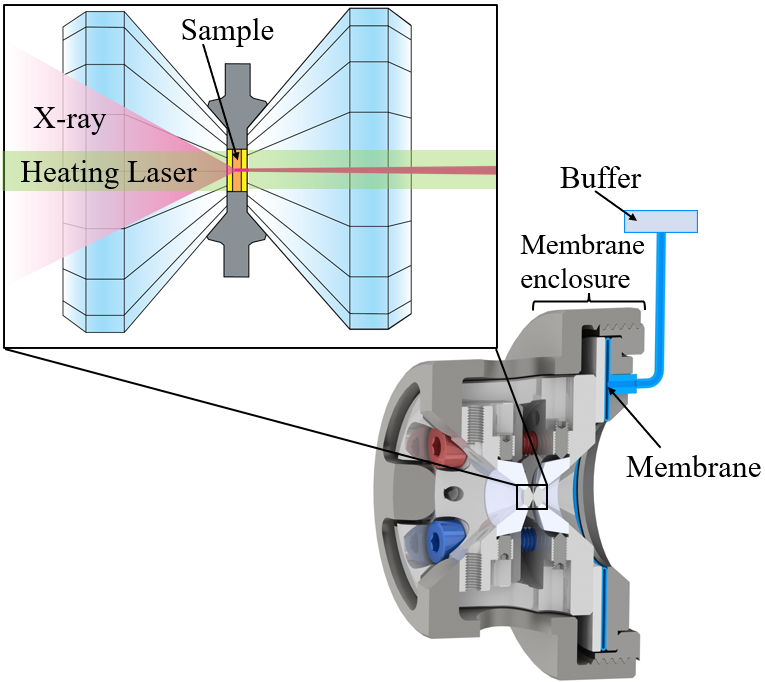}
\caption{Schematic view of the experimental setup used on the 13-IDD beamline at the APS synchrotron. Dynamic loading of the DAC was performed using a membrane and an enclosure compatible with the mini-BX80 cells; the intermediate buffer allowed to perform fast (ms) compression runs. The structure of the sample was monitored in real-time using XRD. The X-ray beam was spatially overlapped with the laser-heating spot and had a wavelength of 0.3344$\AA$.}
\end{figure}

\begin{figure}
\includegraphics[width=0.9\columnwidth]{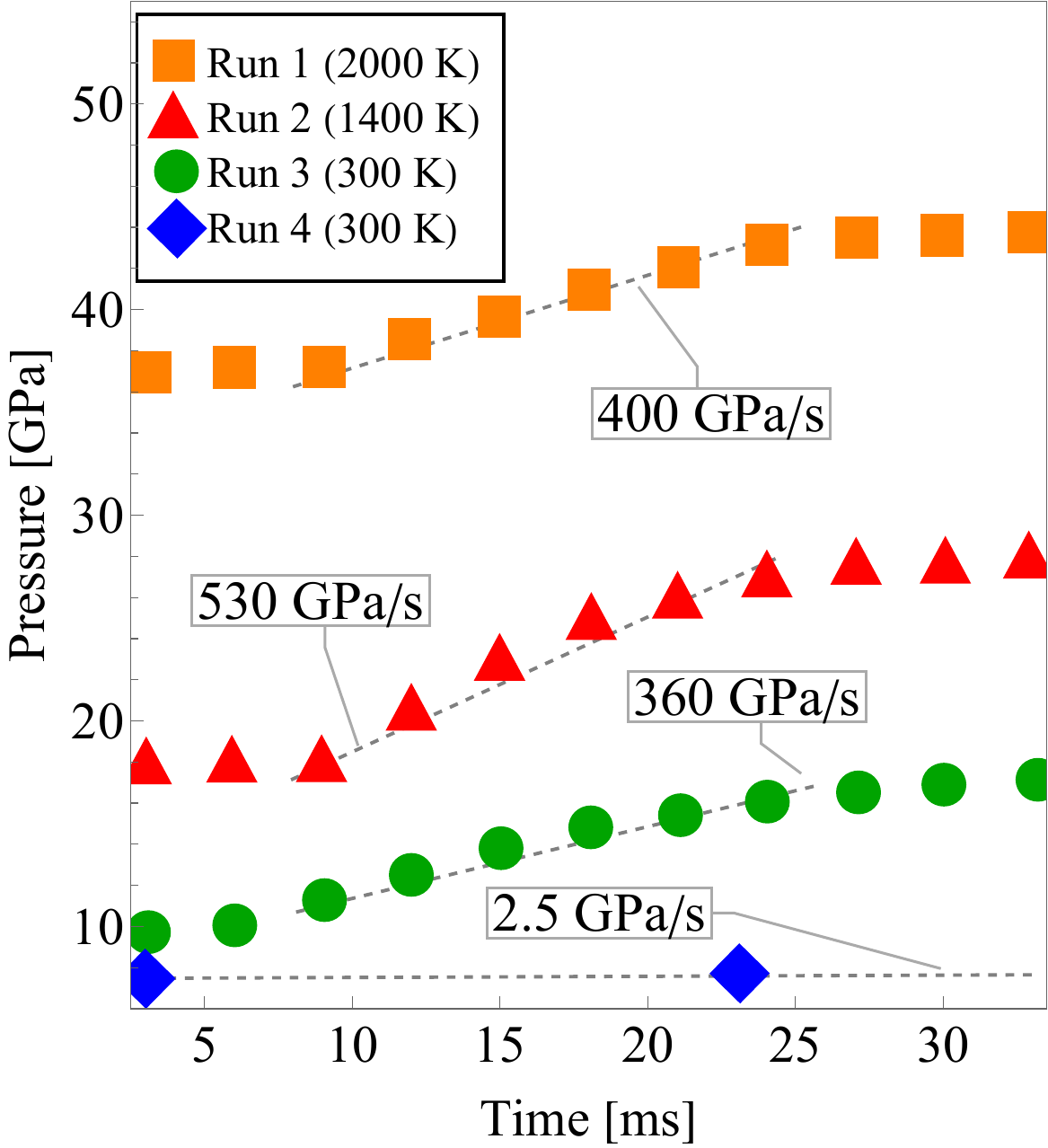}
\caption{\label{fig:comp_rate} Temporal evolution of pressure during dynamic loading for the four experimental runs; pressure was measured using the known EOS of KCl. For each run, temperature values, as well as compression rates, are reported; in Run 4, no intermediate buffer was used, resulting in a $\sim$100 times lower compression rate.}
\end{figure}

\begin{figure*}
\includegraphics[width=0.7\textwidth]{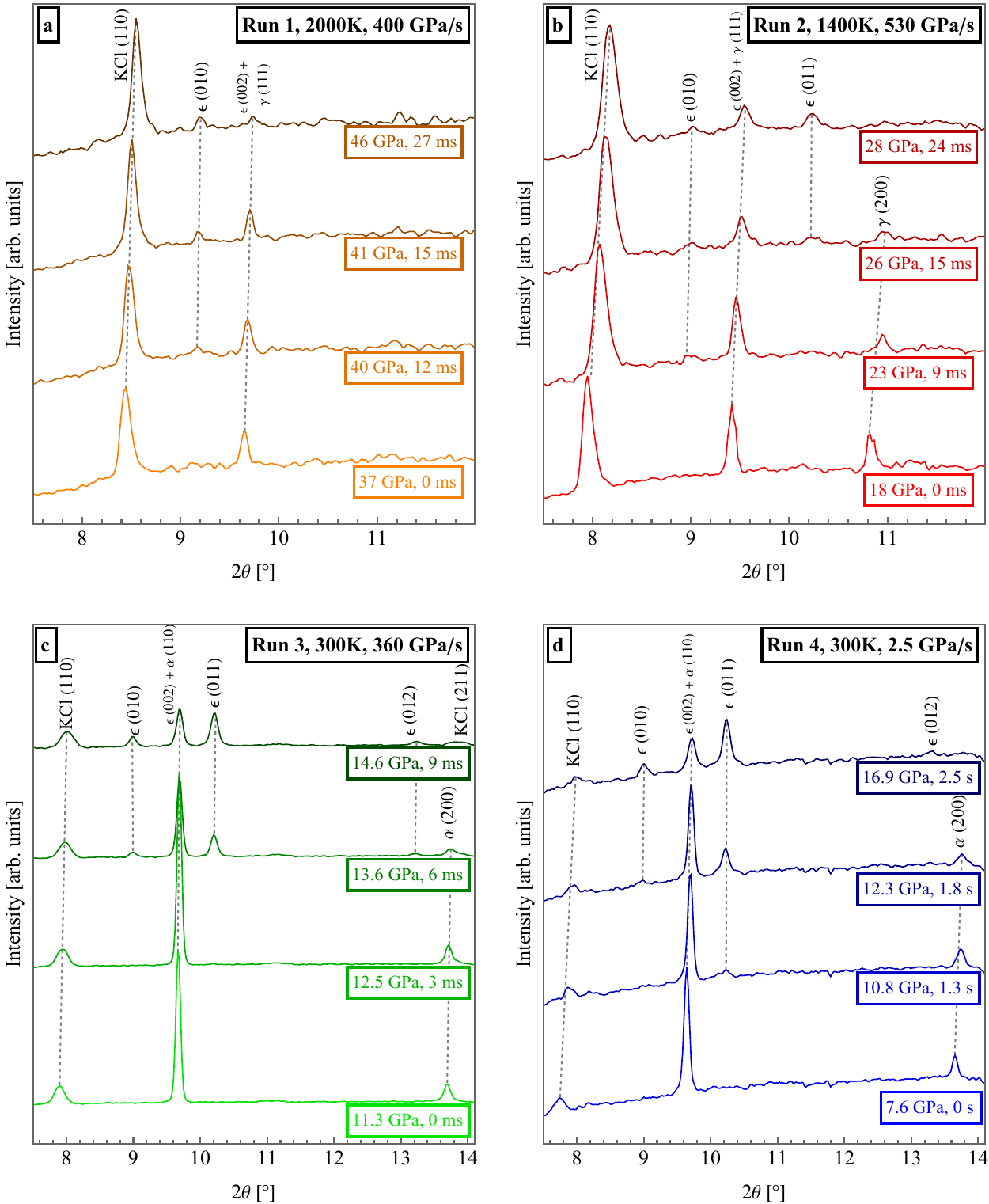}
\caption{\label{fig:waterfall_plots} 
Azimuthally-integrated XRD patterns as a function of time show the structural changes in the sample upon compression. The XRD data is shown in the same colors used in FIG.\ref{fig:comp_rate} to represent pressure evolution. (a) and (b): HT experiments, showing the $\gamma-\epsilon$ transition at 2000 K and 1400 K, respectively. (c) and (d): experiments at 300 K, showing the $\alpha-\epsilon$ transition for compression at 360 GPa/s and 2.5 GPa/s, respectively. For all patterns, the peaks of the observed Fe phases, as well as those of KCl, are indexed; the time values are measured with respect to the beginning of the dDAC compression run.}
\end{figure*}

Experiments were conducted at the 13-IDD beamline of the GSECARS sector of the Advanced Photon Source \cite{Shen2005}. Compression was performed in a mini-BX80 DAC equipped with a membrane enclosure for remote control of the pressure and with an intermediate buffer for fast loading \cite{sinogeikin2015}. The short working distance of the apparatus is compatible with the double-sided laser heating setup of the 13-IDD beamline \cite{prakapenka2008}, and allowed for isothermal dynamic compression runs at HT (see also Sect. I of Supplemental Material). When the intermediate buffer was used, the pressure increased rapidly over few tens of milliseconds, and then kept increasing at a lower rate. Data discussed in the following were acquired during the initial rapid compression, which always comprised the onset of the phase transition; compression rates were estimated via linear fitting of the P evolution over time in the examined interval (FIG. \ref{fig:comp_rate}). In Run 1, compression was performed at 2000 K, with an average compression rate of 400 GPa/s: Run 2 reached 530 GPa/s at 1400 K. Run 3 and Run 4 were performed at 300 K to characterize the $\alpha-\epsilon$ transition reaching 360 GPa/s and 2.5 GPa/s compression rates, respectively. It should be noted that in Run 4 the compression was performed without using the intermediate buffer. X-ray diffraction data were collected \textit{in situ} during the dynamic compression ramp with sufficient temporal resolution for characterization of the structural evolution and accurate determination of the transition onset. During Run 1-3, XRD data were acquired at 2 ms/pattern, while the acquisition time was increased to 20 ms/pattern during Run 4. For each experiment, the sample was compressed statically up to a pressure value close to the phase boundary before launching the dynamic compression run, as to ensure completion of the transformation during loading. 

In Run 1 measurements were performed keeping temperature at 2000 K, and the sample was pre-compressed up to 37 GPa (FIG.\ref{fig:waterfall_plots}(a)); at these conditions only the $\gamma$ phase is present. During loading, we observe the appearance of the $\epsilon$(010) reflection ($\sim$9.2$^\circ$) after 12 ms, at 40 GPa. Coexistence of the $\gamma$ and $\epsilon$ phases is observed up to 46 GPa, and progression of the transformation is confirmed by the changes in relative intensity of the correspondent peaks. At 46 GPa, the peak observed at $\sim$9.7$^\circ$ can be indexed as either the $\gamma$(111) or the $\epsilon$(002) reflections. The crystalline texture, which can be inferred by examining the intensity distribution along the Debye-Scherrer ring in the 2D XRD data, is more consistent with the $\gamma$ phase (see also Sect. III-C of Supplemental Information). Moreover, based on previous experiments, the $\epsilon$ phase is expected to grow along a preferred orientation in a DAC, which results in a decrease of the intensity of $\epsilon$(002) reflection \cite{Singh2006}; thus, we do not expect to observe signal from the $\epsilon$(002) peak over the 2 ms integration time. In Run 2 the sample was pre-compressed up to 18 GPa and temperature was maintained at 1400 K during dynamic loading (FIG.\ref{fig:waterfall_plots}(b)). At $t = 0$ ms, all the Fe peaks can be indexed as reflections from the $\gamma$ phase. The $\epsilon$(010) peak becomes visible after 9 ms, at 23 GPa, and at 28 GPa, after 15 ms, the $\gamma$(200) reflection is no longer visible, leaving only the $\gamma$(111) peak overlapping with the $\epsilon$(002) peak in the diffraction pattern. In Run 3 the sample was initially compressed to about 11 GPa, a pressure at which only the $\alpha$ phase is present. Dynamic loading was performed at 300 K and 360 GPa/s (FIG.\ref{fig:waterfall_plots}(c)); at 13.6 GPa, multiple reflections from the $\epsilon$ phase appear, namely $\epsilon$(010), $\epsilon$(011) and $\epsilon$(012). In Run 4 compression was performed at 300 K without the use of the intermediate buffer, resulting in a compression rate reduction by a factor $\sim$100 to 2.5 GPa/s (FIG.\ref{fig:waterfall_plots}(d)). The sample was compressed statically up to 7 GPa, and at this compression rate the emergence of the $\epsilon$(010) and $\epsilon$(011) reflections was observed at pressures as low as 10.8 GPa, a value that is in agreement with the $\alpha-\epsilon$ phase boundary identified in static compression experiments \cite{Shen1998}. It is worth noting that for all runs, despite persistence of small fractions of the initial phase during dynamic loading, longer-acquisition XRD collected at the end of each compression ramp confirmed that the sample had fully transformed into the $\epsilon$ phase (see also Sect. III-C of Supplemental Material).

\begin{figure}
\includegraphics[width=\columnwidth]{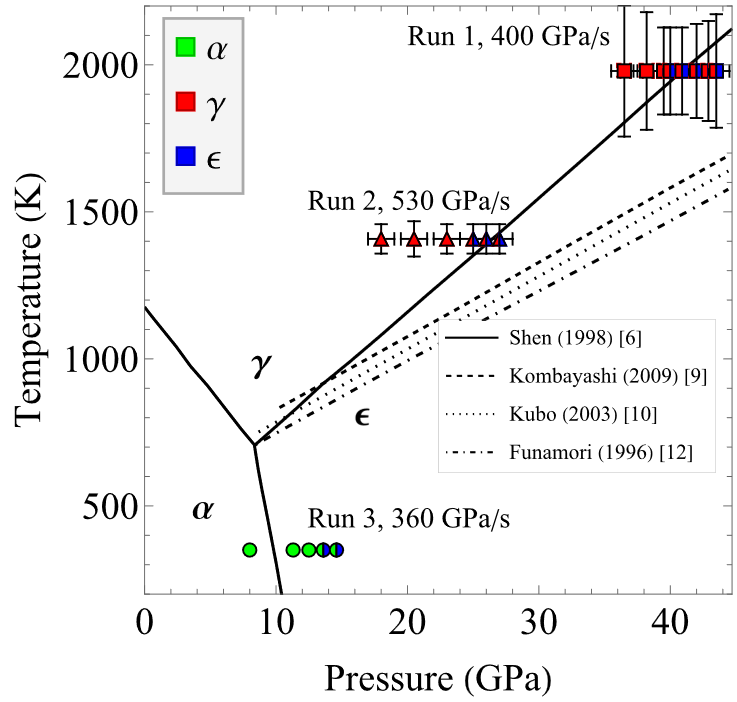}
\caption{\label{fig:phase_dia} Experimental results from dDAC experiments compared with the state-of-the-art equilibrium phase diagram of Fe. Data are represented using markers of different shapes for each run, while the colours correspond to different Fe phases. The solid line is the equilibrium phase diagram by Shen 
\textit{et al.} \cite{Shen1998}; the dotted lines represent the $\gamma-\epsilon$ equilibrium boundary from other experimental studies \cite{Komabayashi2009, Kubo2003, Funamori1996}}
\end{figure}

The experiments here presented use a dDAC apparatus to perform HP-HT experiments and reach compression rates up to $\sim$500 GPa/s to investigate the influence of the loading timescale on Fe behaviour at extreme conditions. The transition onsets measured at different temperatures for both the \textit{fcc}-\textit{hcp} ($\gamma-\epsilon$) and the \textit{bcc}-\textit{hcp} ($\alpha-\epsilon$) phase transitions of Fe are shown in FIG.\ref{fig:phase_dia} and overlaid with previous results from static compression experiments  \cite{Shen1998,Komabayashi2009,Kubo2003,Funamori1996}. For the $\gamma-\epsilon$ transition, results from dDAC are in very good agreement with the phase boundaries proposed by Shen \textit{et al.} \cite{Shen1998}, while they exhibit a lower transition onset compared with more recent static compression experiments \cite{Komabayashi2009, Kubo2003, Funamori1996}. It is worth noting that, independently of the considered reference for the equilibrium phase boundary, no increase in the phase transition onset is observed under dynamic loading. In contrast, for a compression rate of 360 GPa/s at 300 K, the $\alpha-\epsilon$ transition is observed at 13.6 GPa, a higher pressure with respect to the boundary from static compression experiments. To confirm that the shift observed in the $\alpha-\epsilon$ transition is due to the compression timescale, we have performed an additional experiment (Run 4, not shown in FIG.\ref{fig:phase_dia}) at $\sim$100 lower compression rate; the XRD data confirms that, at 2.5 GPa/s the transition onset is 10.8 GPa, in much closer agreement with the equilibrium value. This finding is also consistent with recent laser-driven ramp compression experiments that reported a shift of the $\alpha$-$\epsilon$ transition onset with increasing strain rate \cite{Amadou2016}. Our results demonstrate the capability of the LH-dDAC apparatus to generate high pressures over timescales at which kinetic effects can be observed and characterized.

The influence of strain rate on HP phase transitions has already been analyzed in several systems; however, the effects of fast compression on the phase boundaries strongly depend on the system and on the specifics of the deformation mechanism. For example, in certain systems higher compression rates can cause an increase of the transition pressure, as the fast loading hinders the rearrangement of the atoms (so-called \textit{kinetic hindrance}) \cite{Husband123}. In contrast, several systems have been observed to exhibit a phase transition lowering under dynamic compression, \textit{e.g.}, silicon \cite{McBride2019, Pandolfi2022}, bismuth \cite{pepin2019} and antimony \cite{Coleman2019}. Interestingly, with the exception of silicon, the lowering in pressure is observed for \textit{displacive} transitions, \textit{i.e.}, transformations that require no change in unit cell volume and happen via small displacements of the atoms.

Our results indicate two distinct trends in the $\alpha-\epsilon$ and $\gamma-\epsilon$ phase transitions of Fe under dynamic loading: compared to static compression experiments, the onset of $\epsilon$ formation is increased and unchanged (or lowered), respectively. Based on previous results from dynamic compression experiments, this is likely due to differences in the transition mechanisms that govern the transformations at the atomic level. The \textit{bcc}-\textit{hcp} ($\gamma-\epsilon$) transformation happens via a combination of compression along one axis and shuffle of the planes \cite{Hawreliak2006, Kalantar2005}; recent experiments have confirmed that the completion of the transformation requires two steps: a displacive seeding followed by a \textit{reconstructive} (\textit{i.e.}, involving bond breaking) deformation \cite{BOULARD202030}. Under dynamic compression, a reconstructive transformation is expected to exhibit kinetic hindrance, as also suggested by molecular dynamics simulations of the $\alpha-\epsilon$ transition \cite{shao2018}. On the other hand, the \textit{fcc} and \textit{hcp} structures are more closely related, as they only differ in the stacking of the planes along one direction; the transformation is thus expected to happen via a purely displacive deformation \cite{Olson1976}. Thus, for the \textit{fcc}-\textit{hcp} transition, an increase of the compression rate would induce a high density of stacking faults, which could result in a high number of nucleation sites and ultimately favor the phase transition \cite{pepin2019, Ling2022, Sharma2020, SharmaPRL2020, SharmaPRX2020}. It is also worth noting that, at the strain-rates of our experiments, Fe plastic deformation is predominantly driven by thermally-activated dislocation flow \cite{Smith2011}, so the generation and diffusion of crystalline defects could strongly influence structural transformations in our HP-HT dDAC experiments.

In this study, we demonstrate for the first time, dynamic compression of a material in a dDAC setup coupled with stable laser-heating. Compression rates of hundreds of GPa/s were attained while simultaneously maintaining  high temperatures up to 2000 K. Collection of time-resolved XRD data with millisecond time resolution enabled characterization of the phase transitions of Fe \textit{in situ}. Interestingly, the dDAC laser-heating setup allows exploration of (quasi)isothermal compression of a material, a pathway not attainable using conventional shock-compression techniques. This demonstrates a new approach for exploration of HP-HT phase transitions under dynamic loading, covering an intermediate timescale between the well-established static- and shock- compression methods. With this approach, we provide the first insight on the $\gamma-\epsilon$ phase transition of Fe under dynamic compression, and compare our results with those obtained for the $\alpha-\epsilon$ transition, as well as the equilibrium phase diagram. We observe that the increase in strain-rate affects the phase transitions of Fe differently, and we attribute the differences to the specific deformation mechanisms. Indeed, no sign of kinetic hindrance is observed for the displacive $\gamma-\epsilon$ phase transition up to 500 GPa/s. In contrast, the reconstructive $\alpha-\epsilon$ transition exhibits a marked increase of the transition onset with the compression rate. Thus, particular care should be taken when using experimental data to model extreme conditions processes at different timescales. This study demonstrates a new approach for exploration of HP-HT phase transitions under dynamic loading, covering an intermediate timescale between the well-established static- and shock- compression methods.

\begin{acknowledgments}
This work was carried out at the GeoSoilEnviroCARS (The University of Chicago,
Sector 13), Advanced Photon Source (Argonne National Laboratory). GeoSoilEnviroCARS is
supported by the National Science Foundation (NSF)—Earth Sciences (No. EAR-1634415). The
Advanced Photon Source is a U.S. Department of Energy (DOE) Office of Science User Facility operated for the DOE Office of Science by Argonne National Laboratory under Contract No. DE-AC02-06CH11357. A.E.G., R.L.S., and S.P. acknowledge support from 2019 DOE FES ECA. A.E.G. and W.L.M. acknowledge support by the Geophysics Program at NSF (EAR2049620). M.R. acknowledges support from the College of Physical and Mathematical Sciences at Brigham Young University and DOE SULI 2021 at SLAC National Accelerator Laboratory.
\end{acknowledgments}

\nocite{*}

\bibliography{bibtexfile.bib}

\end{document}